\documentclass[prl,twocolumn,showpacs,floatfix,preprintnumbers,amsmath,amssymb,superscriptaddress]{revtex4}

\usepackage{graphicx}
\usepackage{dcolumn}
\usepackage{bm}
\usepackage{color}
\usepackage{color}
\usepackage[latin1]{inputenc}
\usepackage[urlcolor=blue]{hyperref}
\hypersetup{backref, colorlinks=true, linkcolor=blue, citecolor=blue}

\begin{document}

\title{{High Pressure Behaviour of Superconducting Boron-doped Diamond}}

\author{M Abdel-Hafiez}
\affiliation{Center for High Pressure Science and Technology Advanced Research, Beijing, 100094, China}

\author{Dinesh Kumar}
\affiliation{Department of Physics, Nano Functional Materials Technology Centre and Materials Science Research Centre, Indian Institute of Technology Madras, Chennai, 600036, Tamil Nadu, India}

\author{R. Thiyagarajan}
\affiliation{Center for High Pressure Science and Technology Advanced Research, Shanghai, 201203, China}

\author{Q. Zhang}
\affiliation{Center for High Pressure Science and Technology Advanced Research, Shanghai, 201203, China}

\author{R. T. Howie}
\affiliation{Center for High Pressure Science and Technology Advanced Research, Shanghai, 201203, China}

\author{K. Sethupathi}
\affiliation{Department of Physics, Low temperature physics laboratory, Indian Institute of  Technology (IIT)
Madras, Chennai, 600036, Tamil Nadu, India}

\author{O. Volkova}
\affiliation{Low Temperature Physics and Superconductivity Department, Physics Faculty, M.V. Lomonosov Moscow State University, 119991 Moscow, Russia}
\affiliation{Theoretical Physics and Applied Mathematics Department, Ural Federal University, 620002 Ekaterinburg, Russia}
\affiliation{National University of Science and Technology "MISiS", Moscow 119049, Russia}

\author{A. Vasiliev}
\affiliation{Low Temperature Physics and Superconductivity Department, Physics Faculty, M.V. Lomonosov Moscow State University, 119991 Moscow, Russia}
\affiliation{Theoretical Physics and Applied Mathematics Department, Ural Federal University, 620002 Ekaterinburg, Russia}
\affiliation{National University of Science and Technology "MISiS", Moscow 119049, Russia}

\author{W. Yang}
\affiliation{Center for High Pressure Science and Technology Advanced Research, Shanghai, 201203, China}

\author{H. K. Mao}
\affiliation{Center for High Pressure Science and Technology Advanced Research, Shanghai, 201203, China}

\author{M. S. Ramachandra Rao}
\affiliation{Department of Physics, Nano Functional Materials Technology Centre and Materials Science Research Centre, Indian Institute of Technology Madras, Chennai, 600036, Tamil Nadu, India}

\date{\today}

\begin{abstract}
This work investigates the high-pressure structure of freestanding superconducting ($T_{c}$ = 4.3\,K) boron doped diamond (BDD) and how it affects the electronic and vibrational properties using Raman spectroscopy and x-ray diffraction in the 0-30\,GPa range. High-pressure Raman scattering experiments revealed an abrupt change in the linear pressure coefficients and the grain boundary components undergo an irreversible phase change at 14\,GPa. We show that the blue shift in the pressure-dependent vibrational modes correlates with the negative pressure coefficient of $T_{c}$ in BDD. The analysis of x-ray diffraction data determines the equation of state of the BDD film, revealing a high bulk modulus of  $B_{0}$=510$\pm$28\,GPa. The comparative analysis of high-pressure data clarified that the sp$^{2}$ carbons in the grain boundaries transform into hexagonal diamond.
\end{abstract}


\maketitle
\section{I. Introduction}

Diamond exhibits complex electronic transformations as boron concentration is raised high enough to drive an otherwise insulating system to a metallic regime. It is here, that merging of the impurity and valence bands results in superconductivity in diamond~\cite{blase2009superconducting}. Over a decade of research has shown a steady improvement in the diamond superconducting transition temperature $T_{c}$, from an initial report~\cite{ekimov2004superconductivity} of 2.3\,K to 10\,K, as reported in the most recent work~\cite{Okazaki2015}. More interestingly, Mousa {\it et al.} report that the $T_{c}$ in diamond can be raised up to 55\,K with efficient doping~\cite{Moussa2008}. However, for all practical purposes, beyond a certain boron concentration ($>$ 10$^{19}$\,cm$^{-3}$) the formation of a multiboron complex, segregation of boron, vacancies and interstitial boron have impeded an increase in the active carrier concentration required to achieve the above predicted $T_{c}$~\cite{Goss2006}.  In fact, in polycrystalline boron doped diamond (BDD), it is estimated that only 10\% of the incorporated boron atoms are isolated substitutional boron atoms, whereas the remaining boron atoms are either consumed by the grain boundaries or become point defects that are inactive or even detrimental to the $T_{c}$ ~\cite{chen1999defect,dubrovinskaia2008insight}.

Despite great efforts in the study of the group-IV covalent semiconductors, many unresolved questions and unexplained results require further investigation. These include establishing the nature of their superconducting coupling mechanism~\cite{dubrovinskaia2008insight,marevs2007unconventional,boeri2004three,blase2004role,baskaran2008resonating}. Discovery of superconductivity in diamond was immediately followed by theoretical works from various groups stressing on the study of the mechanism of superconductivity in BDD. \textit{Ab-initio} calculations ~\cite{boeri2004three,blase2004role} have revealed that beyond the critical concentration $n_c = 4.5\times10^{20} \rm{cm}^{-3}$~\cite{klein2007metal}, an insulator to metal transition (IMT) sets in and the impurity band merges with the valence band driving the Fermi level into the valence band. For such degenerate semiconductors, it becomes energetically favourable for the Cooper pair formation via the coupling of holes with the zone centre phonon (ZCP). If $\lambda_{el-ph}$ is the electron-phonon coupling constant and $\omega$ is the ZCP frequency then, BCS theory estimates the superconducting transition temperature $T_c$, using $T_c\sim \omega\rm{ exp{\big(\frac{-1}{\lambda_{el-ph}}\big)}}$. Contrary to the phonon mediated pairing mechanism, an alternative theory proposed by Baskaran suggests the existence of rigid impurity band in the valence band of superconducting BDD, where the width of the impurity band provides an estimate of $T_c$~\cite{baskaran2008resonating,baskaran2006strongly}.  The present experimental results appear to favor the BCS mechanism and this finding could be emphasized later in the paper.
In addition, several questions can be raised; in particular, does granular superconducting BDD undergo a phase change at high pressure$?$ How compressible is  superconducting diamond$?$ More generally, high-pressure investigations in pure diamond have undoubtedly established that cubic diamond is highly incompressible and it retains its cubic structure even up to 140\,GPa~\cite{Occelli2006}. On the other hand, graphite undergoes a phase transformation at 14\,GPa~\cite{Halfhand1989}. This is because the covalently bonded hexagonal planes in graphite are connected by weak van der Waal's bonds that can be easily deformed. Due to puckering of its hexagonal planes, sp$^2$ hybridized carbon atoms transform into sp$^3$ hybridized carbon atoms. The general consensus is that graphite transforms into lonsdaleite at 14\,GPa, also known as hexagonal diamond, with relatively large lattice parameters of $a$ = 2.52\,$\rm{\AA}$ and $c$ = 4.12\,$\rm{\AA}$ compared to $a$ = 1.54\,$\rm{\AA}$, for its cubic counterpart~\cite{Halfhand1989,Brillante1986,Salehpour1990}. As granular BDD consists of both sp$^2$- and sp$^3$-hybridized networks, it can be regarded an ideal system to manifest the above mentioned changes under high pressure. Most of the work after the discovery of superconductivity in BDD is based on the more relevant sp$^3$ networks in the grains and therefore, the present state of knowledge on how boron atoms are accommodated in the sp$^2$ matrix is vague\cite{dubrovinskaia2008insight,lu2015boron,turner2012local,lu2013local,lu2012direct,zhang2014global,mortet2017intrinsic}. Finding a phase change in the heavily doped sp$^{2}$ and sp$^3$ hybridized network at high pressures would solve a longstanding mystery.

To answer these intriguing questions about the phase change and compressibility of granular superconducting BDD and also to shed light on its origin of superconductivity, we deposited a freestanding 60\,$\mu$m thick BDD film using a challenging hot filament chemical deposition (HFCVD) technique. This method can achieve a high $T_{c}$ with sufficient grain boundary content that could reveal phase changes at high pressure and provides an opportunity to explore the relevance of phonon-mediated mechanisms in BDD.

\section{II. Experimental details}

BDD film was grown using hot filament chemical vapour deposition (HFCVD) technique. Prior to the deposition, the silicon substrate was seeded with commercially obtained nano-diamond solution immersed in Dimethyl sulfoxide solution. After loading the pre-treated substrate, the chamber was evacuated to a base pressure of 10$^{-3}$\,Torr. The filaments are heated to 2200\,$^{\circ}$C by passing high current across its ends. Si substrate is placed at a suitable distance away from the filaments such that the temperature around it is 850\,$^{\circ}$C.  Deposition was carried out by maintaining the chamber pressure at 7\,Torr. CH$_{4}$, H$_{2}$ and (CH$_3$)$_3$B flowrates were maintained at 80 sccm, 3000 sccm and 35 sccm, respectively.  Further details on the HFCVD reactor can be found elsewhere~\cite{chandran2012integration}. After the deposition for 110 h, the Si substrate was etched away using KOH solution leaving behind a free standing 60 $\mu$m thick BDD film with grain sizes $<$ 1 $\mu$m. Surface morphology and the thickness of the BDD sample was determined using SEM (QUANTA 3D FEG microscope).

Electrical transport, specific heat measurements, and magnetic measurements were carried out using the physical property measurement system (PPMS). Resistance measurements down to 100\,mK were performed using dilution refrigerator. Four contacts were used to measure the high-pressure resistivity. The hydrostatic pressure up to 30\,GPa was generated using a symmetrical diamond anvil cell device (DAC) employing diamond anvils with a culet size of 300\,$\mu$m with silicon oil as a pressure medium. The sample of size 50\,$\mu$m was placed in a 150\,$\mu$m diameter hole on a SS T301 gasket. The laser wavelength used for the Raman measurement was 632.8\,nm He - Ne laser using gratings of 1800\,gr/mm. High-pressure Raman spectra were carried out at HPSTAR, China. The {\it in-situ} high-pressure X-ray diffraction experiments were carried out using the synchrotron facility at the High-Pressure Collaborative Access Team (HPCAT) at the Advanced Phonon Source (APS), USA, with a wavelength of 0.3100\,$Å$. The diffraction data were recorded with two-dimensional (2D) images, then Dioptas was used for integration, and the structure was refined to analyze the XRD data. For the \textit{in-situ} high-pressure studies, the pressures were determined by the ruby fluorescence method.  Ruby chips were put near the crystal and NaCl powders were dropped surrounding the crystal to serve as pressure transmitting medium. The pressure was monitored by the ruby fluorescence~\cite{Abdel2016}.

\section{III. Results and discussions}

\subsubsection{A. Thermodynamic measurements
}

We ascertained the bulk superconductivity of a 60\,$\mu$m thick granular superconducting BDD film using electrical transport, magnetic and specific heat studies. The resistivity versus temperature curve of the thick BDD film presented in Fig.\,1(a) shows the sample's weak semiconducting behaviour at higher temperatures followed by a sharp drop in resistivity with an offset superconducting transition at $T_{c}$ = 4.3\,K. This can also be seen by the diamagnetic response of the BDD film in Fig.\,1(d).

\begin{figure*}[tbp]
\includegraphics[width=37pc,clip]{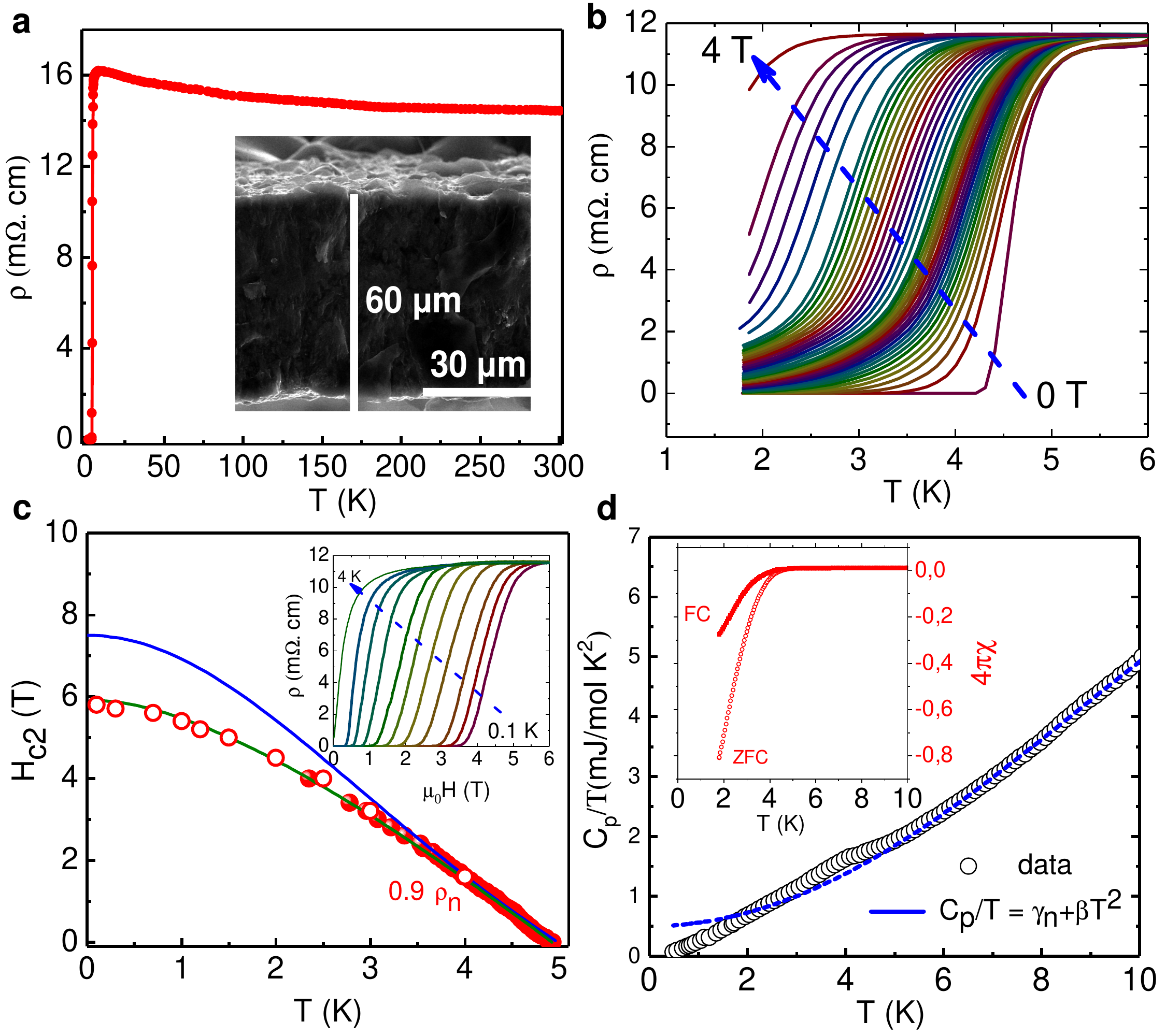}
\caption{\label{fig:wide}  (a) Resistivity versus temperature curve for the BDD film with boron concentration $n_{B}$ = 2.7 $\times$ $10^{21}$\,cm$^{-3}$, as estimated using Raman spectroscopy~\cite{Bernard2004nondestructive,prawer2004raman,may2008raman}. Inset: cross-sectional SEM image of the 60\,$\mu$m thick BDD film. (b) and the inset of (c) present temperature and field dependence down to 100 mK of transport measurements of 60\,$\mu$m thick BDD film. (c) Illustrates the upper critical field $H_{c2}$ extracted using the 90\%$\rho _{n}$ criteria. Open symbols are taken from the resistivity versus magnetic field curves. (d) Temperature dependence of the total specific heat in zero magnetic field. The solid line is the specific heat fitting below 10\,K using $C_{p}/T$ = $\gamma_n$ + $\beta T^{2}$, here, the inset illustrates temperature dependence of the magnetic susceptibility $\chi$, in an external field of 10\,Oe.  $\chi$ was deduced from the dc magnetization, measured by following the ZFC and FC protocols.}
\end{figure*}

\begin{figure*}[tbp]
\includegraphics[width=43pc,clip]{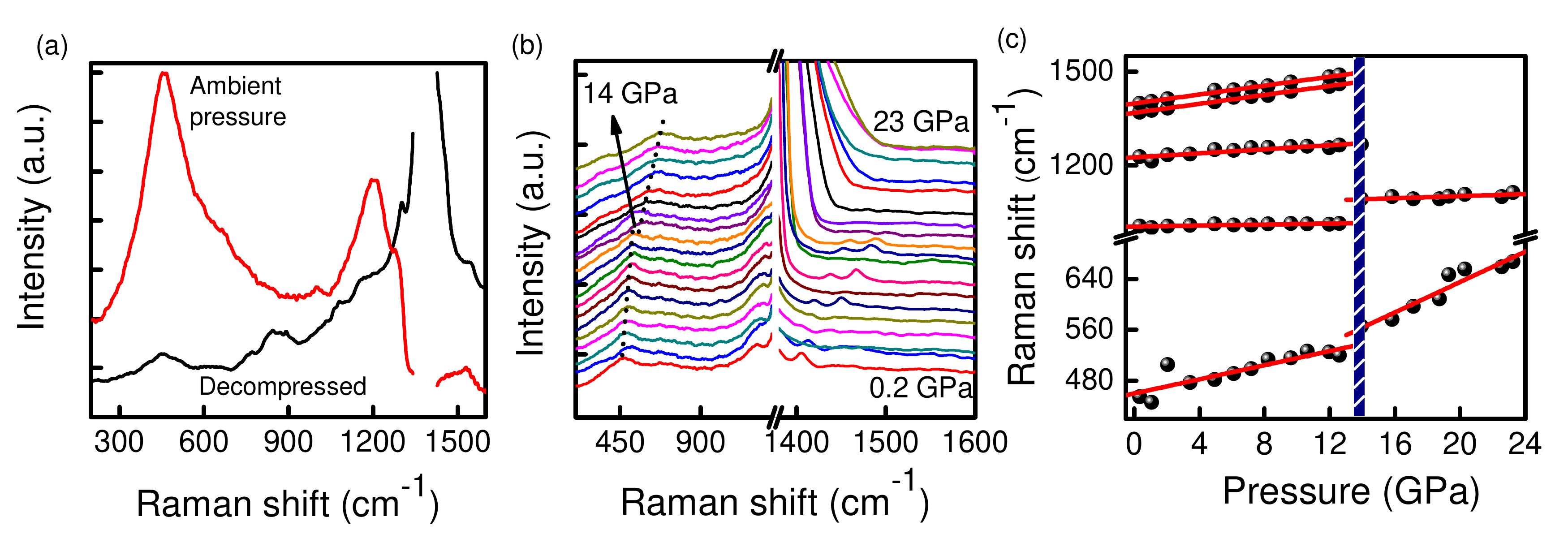}
\caption{\label{fig:wide}  (a) Raman spectra of BDD at ambient pressure (red) and the decompressed Raman spectrum at 0.2\,GPa (black). (b) Raman spectra of freestanding BDD under hydrostatic pressure; dotted lines indicate the abrupt change in the pressure coefficient at 14\,GPa. The intense signal at $\sim$1300\,cm$^{-1}$ is due to the diamond anvil. (c) Pressure dependence of various vibrational modes in the superconducting BDD; solid spheres represent the peak centre obtained using a Gaussian fit on various peaks and the solid line represents a linear fit. Here, the hatched line indicates an abrupt change in the pressure coefficient.}
\end{figure*}

The critical temperature dependencies of $H_{c2}$ were extracted using the 90\%$\rho _{n}$ criteria from the magnetoresistance curves to 100\,mK. Here, $\rho _{n}$ represents the resistivity just above the onset $T_c$. Unlike in the case of single crystalline superconducting BDD~\cite{bustarret2004dependence}, Ginzburg-Landau (GL) extrapolation as represented by the blue curve, does not reproduce the data at low temperatures. Hence, we adopted the Werthamer-Helfand-Hohenberg (WHH) model, the curve in olive in Fig.\,1c, to fit the $H_{c2}(T)$ versus $T_{c}$ plot. The WHH theory predicts the behavior of $H_\mathrm{c2}(T)$ in the dirty limit taking into account paramagnetic and orbital pair-breaking~\cite{Werthamer1966}. The temperature dependence of $H_{c2}$ is given by the WHH formula,
\begin{equation}\label{eq3}
    \ln\frac{1}{t}=\sum_{\nu=-\infty}^{\infty}\{\frac{1}{2\nu+1}-[2\nu+1+\frac{\bar{h}}{t}+\frac{(\frac{\alpha\bar{h}}{t})^{2}}{2\nu+1+\frac{\hslash+\lambda_{so}}{t}}]^{-1}\},
\end{equation}
where $t = \dfrac{T}{T_{c}}$, $\hbar = \dfrac{4}{\pi^{2}}{H_{c2}(T)}\left|\dfrac{dH_{c2}}{dT}\right|^{-1}_{Tc}$, $\alpha$ is the Maki parameter which describes the relative strength of orbital breaking and the limit of paramagnetism, and $\lambda$$_{so}$ is the spin-orbit scattering constant. The orbital limited upper critical field $H_{\mathrm{c2}}$ at zero temperature is determined by the slope at $T_\mathrm{c}$ as
$H_{\mathrm{c2}} = 0.69 \, T_{\mathrm{c}}\, \dfrac{\partial H_{\mathrm{c2}}}{\partial T}\bigg|_{T_{\mathrm{c}}}$. The curve of $H_{c2}(T)$ had a slope $-\dfrac{dH_{c2}}{dT}$=2.19\,T/K for 90\%$\rho _{n}$. Thus, a fit to the data in the whole measurement range for negligible spin-paramagnetic effects ($\alpha = 0$) and spin-orbit scattering ($\lambda = 0$) yields $\mu_0 H_{\mathrm{c2}} = 5.9$~T~\cite{Clogston1962,Chandrasekhar1962,Werthamer1966}. The coherence length was estimated using $\xi_{GL}=\left(\frac{\phi_0}{2\pi H_{c2}^{WHH}(0)}\right)^{\frac{1}{2}}$ = 7.4\,nm, where $\phi_0$  is the flux quantum. This is in close agreement with $\xi _{GL}$= 7.1\,nm. and $\xi _{GL}$= 10\,nm, as found by Zhang \textit{et al.}~\cite{zhang2014global} and Ekimov \textit{et al.}~\cite{ekimov2004superconductivity}, respectively. While superconductivity is well distinguished by resistivity and magnetization measurements, we further confirmed the bulk superconductivity by performing low-temperature heat capacity measurements down to 0.4\,K, as illustrated in Fig.\,1d. Specific heat has probably the best energy resolution among all experimental probes for distinguishing the bulk superconductivity~\cite{m0,m1,m2,m3}. Sidorov \textit{et al.} conducted specific heat measurements on high-pressure high-temperature (HPHT) grown BDD, revealing the existence of enormous inhomogeneity in their sample~\cite{sidorov2005superconducting}. We performed measurements on a CVD-grown superconducting BDD, which is known to be macroscopically homogenous~\cite{willems2009intrinsic}. Low-temperature specific heat data are plotted as $C_{p}/T$ versus $T$ in Fig.\,1(d). We observed a clear anomaly at $T_{c}$ = 4.8\,K, close to that determined by our resistivity and magnetization measurements. The zero-field specific heat above $T_{c}$ were well fitted to $C_{p}/T$ = $\gamma_n$ + $\beta T^{2}$, with $\gamma_n$ and $\beta$ the electronic and lattice coefficients, respectively (as indicated by the dashed line in Fig.\,1(d)). We found $\gamma_n$ = 0.5\,mJ/mol K$^{2}$ and Debye temperature was extracted using the relation $\theta_D = (12\pi^{4}  R N/5 \beta)^{1/3}$. We obtained \emph{$\theta_{D}$} = 1410(5)\,K, which is comparable with value (\emph{$\theta_{D}$} = 1440\,K) reported by Sidorov \textit{et al.}. Clearly, this is less than \emph{$\theta_{D}$}( = 1880\,K)~\cite{victor1962heat} for the single crystalline diamond, a result of lattice softening to due heavy boron doping.

\begin{table*}
\caption{\label{tab:table 1}  Summary of the pressure coefficient (cm$^{-1}$GPa$^{-1}$) of various Raman modes in granular BDD. We show the pressure coefficient $\left(\dfrac{dw_{i}}{dp}\right)$ at pressures, $<$14 GPa and at $>$14 GPa.}

\begin{ruledtabular}
\begin{tabular}{cccccc}
Vibrational modes & B-B mode & 1000 cm$^{-1}$ & 1200 cm$^{-1}$ & D-Peak & $\nu _{3}$\\
\hline
\\
$\frac{dw_{i}}{dp}$ ($<$14 GPa) &6.3 &0.8&3.24&7.1&7.1\\
\hline
\\
$\frac{dw_{i}}{dp}$ ($>$14 GPa) &11.9&1.6&Masked by anvil&Phase change &phase change \\

\end{tabular}
\end{ruledtabular}
\end{table*}

\subsubsection{B.  High-pressure Raman spectroscopy}

We first discuss our Raman spectroscopy data which suggest a possible phase change at 14\,GPa. The Raman spectrum recorded at ambient pressure and the decompressed spectrum are presented in Fig.\,2(a). The broad peak at 455\,cm$^{-1}$ is due to the B - B dimer; an A$_{1g}$ stretching mode. The peaks at 1000 cm$^{-1}$ and 1204 cm$^{-1}$ are due to the phonon density of states \textcolor{red}{\cite{boeri2004three,szirmai2012detailed}.} The hump at 1280\,$^{-1}$ is due to the Fano resonance of the ZCP of diamond. The graphitic components, including trans-polyacetylene and the D-peak, have Raman signals in higher wave numbers (1350\,cm$^{-1}$ to 1550\,cm$^{-1}$)\cite{mortet2017intrinsic,bourgeois2006dimer,ferrari2004grain,Bernard2004nondestructive,zinin2006pressure}.  We have omitted the Raman modes around 1300\,cm$^{-1}$ because unfortunately, the strong signals from the diamond anvil masked the signals from BDD. Systematic changes of the BDD Raman bands under hydrostatic pressure up to 23\,GPa are presented in Fig.\,2(b) and the corresponding changes are shown in Fig.\,2(c). The values of the pressure coefficients for various Raman modes are listed in Table 1. All the peaks in the Raman spectra clearly undergo blueshift with hydrostatic pressure increase and the pressure coefficient of the B-B mode was the highest. The higher frequency vibrational modes related to the grain boundaries disintegrated and vanished completely at pressures close to 14\,GPa. The pressure coefficient for all the modes that were available up to 23\,GPa doubled abruptly at 14\,GPa.

\begin{figure*}[tbp]
\includegraphics[width=38pc,clip]{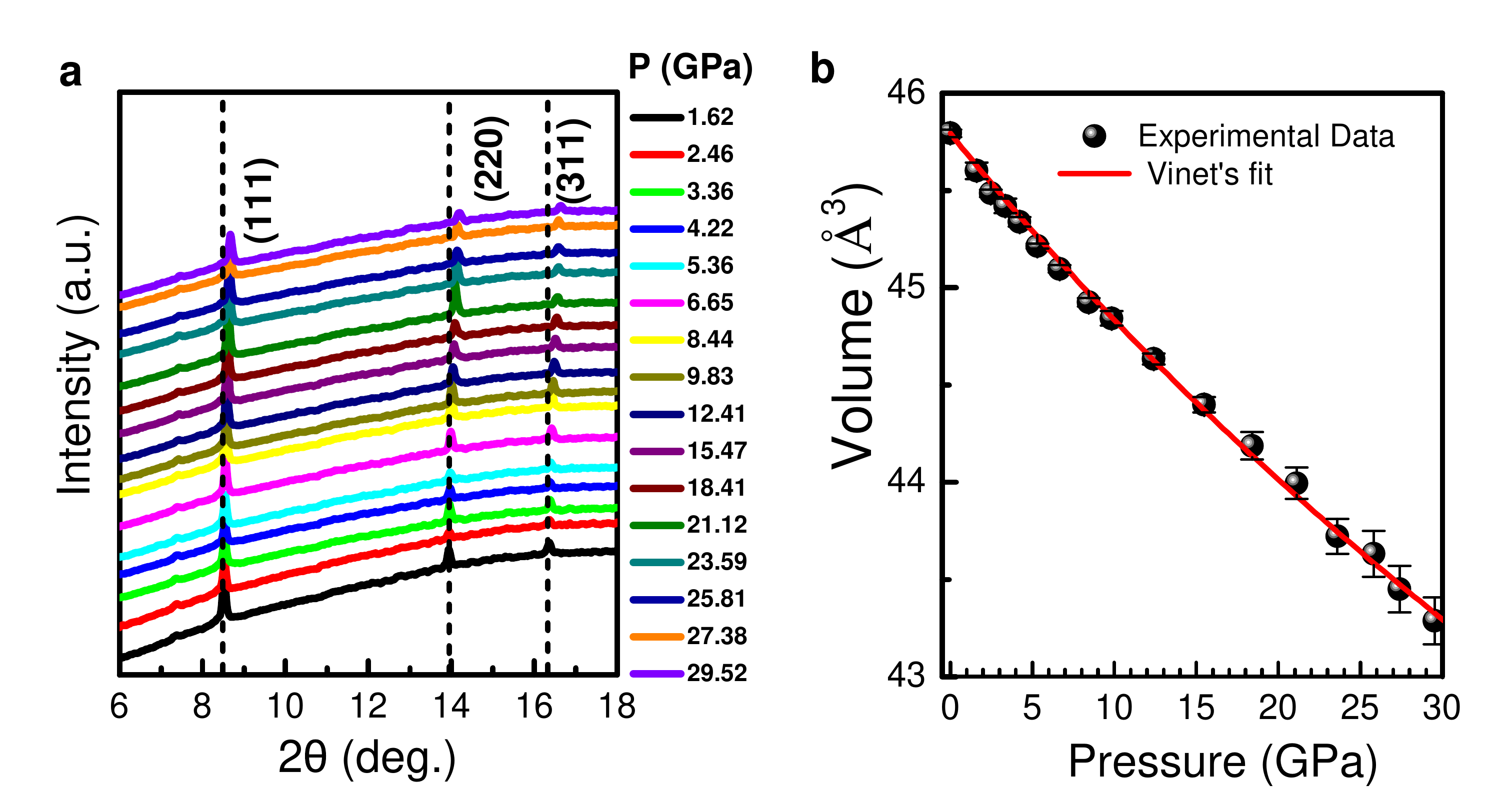}
\caption{\label{fig:wide}  (a) Representative XRD patterns of BDD under various pressures. The dotted lines indicate the shift of the diffraction peaks; a result of lattice contraction under compression. The extra peak at 7.32 degree is due the Fe in the steel gasket. (b) volume change versus the applied pressure behaviour of BDD, where the solid spheres represent the volume at a given pressure and the continuous curve represents a fit to the Vinet's equation.}
\end{figure*}

The blue shift in the Raman modes is due to the frequency dependence of its force constants~\cite{weinstein1984pressure}. Additionally, in BDD the B-B bond length of 1.94 $\rm{\AA}$ is larger than the C-C bond length (1.54 $\rm{\AA}$). This directly infers that B-B bond compression is much easier compared to  C-C bond compression and thus, the pressure coefficient of the B-B mode is relatively high. Close examination of the Raman spectra in the higher frequency region shows various Raman modes originate from the grain boundaries, including the Raman signals at 1370\,cm$^{-1}$ (D peak), 1450\,cm$^{-1}$ (trans-polyacetylene), and the graphitic 1550\,cm$^{-1}$ peak. In our experiment, the Raman bands around 1370\,cm$^{-1}$ and 1550\,cm$^{-1}$ vanished completely above 14\,GPa, suggesting a phase change of the grain boundary sp$^{2}$ complex. The sp$^{2}$ components may have buckled or puckered at relatively low pressure, thereby facilitating a quicker phase change under compression. Such a phase change is a well-known phenomenon in trans-polyacetylene and graphite, where they undergo an irreversible phase change into hexagonal diamond at pressures above 14\,GPa~\cite{Halfhand1989,Brillante1986}. The decompressed Raman data wherein additional peaks are seen at 850\,cm$^{-1}$ and 1300\,cm$^{-1}$. These peaks may occur due to either the increased disorder in the sp$^3$ system or the signals from phase transformed grain boundary carbon network~\cite{knight1989characterization} and more experiments are being carried out to confirm this. The sp$^3$-bonded phase remains intact, as shown by the 500\,cm$^{-1}$ and 1200\,cm$^{-1}$ bands in the decompressed Raman spectrum and we will further validate this using a pressure dependent XRD measurement. Since all the Raman modes in BDD undergo blue  shift it is likely that these modes are stretching modes.

\begin{figure}[b]
\includegraphics[width=19pc,clip]{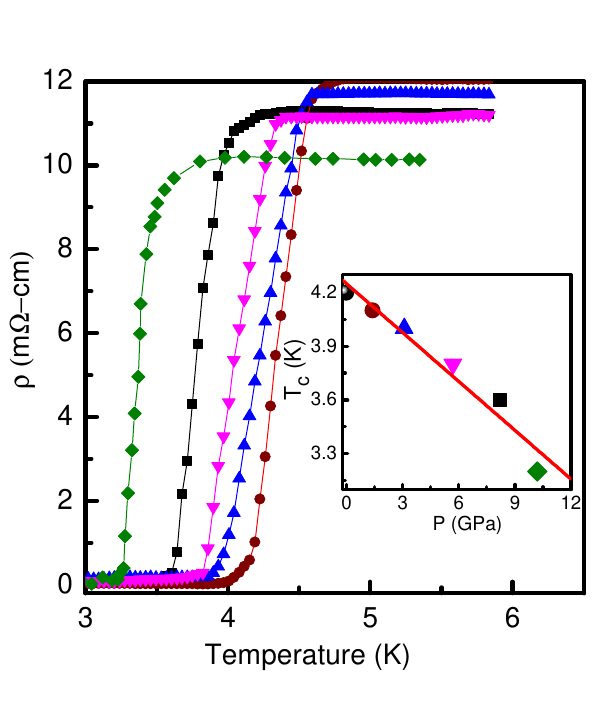}
\caption{Temperature dependence of resistivity under various applied pressures. Superconductivity is suppressed with increasing pressure. Inset: pressure-induced suppression of the $T_{c}$.}
\end{figure}

\subsubsection{C.  High-pressure X-ray diffraction}

On the other hand, properties of the bulk modulus $B_{0}$ and its pressure derivative $B_{0}'=\dfrac{dB_{0}}{dP}$ give a valuable insight into the bonding nature of solids. It is well reported that the cubic diamond phase remains stable up to 140\,GPa and its $B_{0}$ = 442\,GPa~\cite{Occelli2006} and  $B_{0}'$ = 3.6 are well known.  However, similar information for superconducting CVD grown BDD is still missing and needs to be investigated. As the observed changes in the pressure coefficient occur in the Raman modes originating from the sp$^3$ grains, it is tempting to attribute this observation to a phase change in the sp$^3$-bonded grains. The pressure-dependent XRD results in Fig.\,3(a) and 3(b) suggests otherwise. High-pressure XRD results show that the cubic phase of superconducting BDD remains intact up to 30\,GPa. Hence, it is unlikely that the abrupt change in the pressure coefficient is due to a phase change in the sp$^3$-bonded carbons. A plausible explanation for this may be that the doubling of the pressure coefficient is driven by the phase transformation in the grain boundaries. We believe that the pressure coefficient doubling above 14 GPa in the superconducting BDD is due to the conversion of sp$^2$ to sp$^3$-phase in the grain boundaries. Surprisingly, this new phase of sp$^3$-hybridized atoms leads to an increase in the pressure coefficient, defying conventional wisdom that an increase in the coordination number decreases the pressure coefficient~\cite{sherman1980bond,zallen1974pressure}. One plausible explanation for this is that the newly formed sp$^3$ phase above 14\,GPa has larger atom-atom bond lengths~\cite{yagi1992high} than cubic diamond, thereby influencing the ease of compression (or increasing the pressure coefficient) of the whole system.  Figure\,4(d) shows the least square fittings of our P-V data using Vinet's EOS\cite{Vinet1985}, which gives $V_{0}$= 45.70 $Å^{3}$, $B_{0}$= 510\,GPa, and $B_{0}'$  = 2.6. The error bars indicate standard deviations in the estimation of the unit cell volume using the lattice parameters $a_{111}$, $a_{220}$ and $a_{311}$. Unit cell deformation because of random boron incorporation is obvious. However, this deformation becomes noticeably large at higher pressure, as indicated by the error bars. The impressively large $B_0$ in our CVD grown superconducting BDD thick film is noteworthy and although this value is smaller than the optical grade polycrystalline diamond films~\cite{gray1992}, it is certainly larger than HPHT-grown BDD~\cite{ekimov2014synthesis,Dubrovinskaia2006high} or cubic BC$_{5}$~\cite{Solozhenko2009equation}.  $B_0$ depend significantly on the grain sizes~\cite{dubrovinskaia2005aggregated,williams2010high}. It is also possible that, presence of isolated boron rich secondary phase in the $\rm{sp^3}$ bonded grains~\cite{dubrovinskaia2008insight,lu2015boron} may influence the value of $B_0$ in HPHT grown BDD.

Finally, we discuss the relevance of phonon-mediated mechanisms in the grains by conducting high-pressure transport studies. In Fig.\,4, we show our pressure dependent resistivity curves. In our experiment, $T_{c}$ reduces with applied pressure by a pressure coefficient of -0.09\,K/GPa, which is slightly larger than the pressure coefficient of -0.06\,K/GPa, reported by Ekimov \textit{et al.}~\cite{ekimov2004superconductivity}. The suppression of superconductivity at high pressure via a decrease of the electron-phonon coupling parameter in BDD has been theoretically demonstrated~\cite{Ma2005constraints}.  Ma \textit{et al.} pointed out that the decrease in the electronic density of states (N) near the Fermi level and the weakening of $\lambda _{el-ph}$ with increase in pressure causes $T_c$ depression~\cite{Ma2005constraints}. From our result and also in the publication by Tomioka ~\textit{et al.}~\cite{Tomioka2008pressure}, a decrease in $\rho_n$ is observed with increase in pressure. This undermines the possibility of the reduction of N with increase in pressure. Thus, hardening of the Raman ZCP mode, which is well known in the case of intrinsic diamond ~\cite{Occelli2006,Hanfland1985}, however, unfortunately masked by the signals from the DAC in the present experiment, is the most likely reason for the $T_c$ depression under pressure. This has important implications for coupling between the phonon and holes. In the BCS formulation, $\lambda _{el-ph}$= $\frac{ND}{M\omega^{2}}$, where $D$ is the deformation potential and $M$ is the mass of the C atom. However, $\lambda _{el-ph}$ is related to $T_{c}$ as $T_{c}\propto\exp$(-$\frac{1}{\lambda _{el-ph}}$). Therefore, a decrease in $\lambda _{el-ph}$ with pressure results in the reduction of $T_{c}$. Hence, $T_{c}$ suppression with applied pressure can be attributed to hardening of the Raman mode.

\section{IV. Conclusions}

In summary, we used CVD method, one of the most promising ways to grow BDD, to investigate the details of its bonding, vibrational properties, and possible phase changes at high pressures. The grain boundary components undergo an irreversible phase change at 14\,GPa. The pressure coefficient values increased from 6.3\,cm$^{-1}$/GPa and 0.8\,cm$^{-1}$/GPa to 11.9\,cm$^{-1}$/GPa and 1.6\,cm$^{-1}$/GPa, for the Raman modes at 455\,cm$^{-1}$ and 1000\,cm$^{-1}$ respectively. We found a high bulk modulus, $B_{0}$ =510\,GPa in our CVD grown BDD thick film. We show that the blue shift in the pressure-dependent vibrational modes correlates with the negative pressure coefficient of $T_{c}$ in BDD. By comparing our high-pressure XRD and Raman scattering results, we show that the $sp^{2}$ carbons in the grain boundaries transform into hexagonal diamond.

We are grateful to Goran Karapetrov, Jiuhau Chen, Jun Zhao, and Donglai Feng for stimulating discussions. The work in Russia was supported in part from the Ministry of Education and Science of the Russian Federation in the framework of Increase Competitiveness Program of NUST "MISiS" (K2-2015-075) and by Act 211 of the Russian Federation Government, contract: 02.A03.21.0006. DK and MSRR would like to thank the financial support from Department of Science and Technology (DST), New Delhi, that led to the establishment of Nano Functional Materials Technology Centre (NFMTC)(SR$/$NM$/$NAT$/$02$-$2005) and the Department of Atomic Energy (DAE).

\end{document}